\documentclass[a4paper,11pt]{article}
\usepackage{pos}
\usepackage{bm}

\title{A physicist-friendly reformulation of the Atiyah-Patodi-Singer index (on a lattice)}

\author*{Hidenori Fukaya}
\affiliation{Department of Physics, Osaka University\\
  Machikane-yama-cho 1-1, Toyonaka 560-0043, Japan}

\abstract{
  The Atiyah-Singer index theorem on a closed manifold is well understood and appreciated in physics. On the other hand, the Atiyah-Patodi-Singer index, which is an extension to a manifold with boundary, is physicist-unfriendly, in that it is formulated with a nonlocal boundary condition. Recently we proved that the same index as APS is obtained from the domain-wall fermion Dirac operator. Our theorem indicates that the index can be expressed without any nonlocal conditions, in such a physicist-friendly way that application to the lattice gauge theory is straightforward. The domain-wall fermion provides a natural mathematical foundation for understanding the bulk-edge correspondence of the anomaly inflow.\\
  \par
  Preprint number: OU-HET-1124
}

\FullConference{%
 The 38th International Symposium on Lattice Field Theory, LATTICE2021
  26th-30th July, 2021
  Zoom/Gather@Massachusetts Institute of Technology
}

\begin{document}
\maketitle

\section{Introduction}

The Atiyah-Singer (AS) index theorem \cite{Atiyah:1963zz} on a closed manifold without boundary $X$,
\begin{align}
{\rm Ind} (D)= 
\frac{1}{32\pi^2}\int_{X} d^4x \epsilon_{\mu\nu\rho\sigma}{\rm tr}[F^{\mu\nu}F^{\rho\sigma}],
\end{align}
is well-known and appreciated in physics.
The number of chiral zero modes $n_\pm$ with the $\pm$ chirality of the Dirac operator $D$
defines the index ${\rm Ind}(D) = n_+-n_-$, which is related to
the winding number or topological charge of the gauge fields.
The index is essential in understanding
the tunneling effect among different vacuua in quantum-chromo dynamics (QCD).
This text-book-level formula can be confirmed by only one-loop computations,
and therefore, physicist-friendly.

However, the Atiyah-Patodi-Singer (APS) index theorem \cite{Atiyah:1975jf}, which is
an extension of the AS theorem to a manifold $X$ with boundary $Y$,
\begin{align}
  \label{eq:APS}
{\rm Ind}(D_{\rm APS}) = 
\frac{1}{32\pi^2}\int_{X} d^4x \epsilon_{\mu\nu\rho\sigma}{\rm tr}[F^{\mu\nu}F^{\rho\sigma}] \textcolor{black}{-\frac{\eta(iD_Y)}{2}},
\end{align}
is less known. 
Here the second term is called the eta invariant, which is given by a regularized difference
between the number of positive and negative eigenvalues of the boundary Dirac operator $iD_Y$.
This is not surprising since we have been not very interested in manifolds with boundary in physics until very recently.

Recently the theorem became important in condensed matter physics,
as it was pointed out in \cite{Witten:2015aba}
that the APS index is a key to understand bulk-edge correspondence \cite{Hatsugai}
of symmetry protected topological insulator, which is a gapped material in the bulk
but a good conductivity is seen on the edge.
The carrier of the charge, the so-called edge mode,
is described as a massless $2+1$-dimensional  Dirac fermion,
whose partition function suffers from an anomaly of the time-reversal symmetry.
This anomaly is not a problem in the total system since
it is precisely canceled by the bulk fermion determinant.
Each piece of the APS index in Eq.~(\ref{eq:APS}) corresponds
to each phase of the fermion determinants. Namely, the APS theorem
is a mathematical guarantee that the time reversal symmetry
is protected in the total system.
However, the left-hand side of the theorem defined
by a massless Dirac operator with nonlocal boundary condition
is physicist-unfriendly.

The difficulty is in the fact that if we impose local and Lorentz symmetric boundary condition,
the reflected particle at the boundary flips its momentum but keeps the
angular momentum unchanged. This means that the chirality is not conserved
and the $\pm$ chirality sectors do not decouple anymore: $n_\pm$ and the index do not make sense.

The Atiyah-Patodi-Singer boundary condition \cite{Atiyah:1975jf} gives up the locality
and rotational symmetry to respect the chirality.
In the vicinity of the boundary, the Dirac operator has a form
\begin{align}
D = \gamma_4\left(\partial_4 + H \right), 
\end{align}
where we take $x_4$ in the normal direction to the boundary,
the $A_4=0$ gauge is chosen,
and $H = \gamma_4\sum_i \gamma_i D^i $ is a Hermitian operator.
The APS boundary condition is given by this $H$ operator,
such that its positive eigenmode's components are forced to zero at the boundary: $(H+|H|)\psi=0$.
This is a non-local boundary condition requiring
information of eigenfunction extended on the whole boundary manifold.
Note that $H$ commutes with $\gamma_5$ and chirality is conserved.
Then we can define the index by the chiral zero-modes.

This outcome is mathematically beautiful but physicist-unfriendly,
since locality (or causality) is much more important than chirality for physicists,
otherwise, we may allow information to propagate faster than speed of light.
In physics, we should not accept the non-local APS boundary condition no matter how mathematically beautiful it is
\footnote{
  In Refs.\cite{Witten:2019bou, Kobayashi:2021jbn}, it was shown that
  the non-local feature of the APS boundary condition has no problem
  when it is Wick-rotated to a ``state'' at a time-slice.
  }.
Instead, we need to give up the chirality and consider left-right mixing of massive fermion.
The question is if we can make a fermionic integer even within massive systems. 
Our answer is ``Yes we can''.

Here is our reference list. We proposed a new formulation of the APS index using the domain-wall fermion
\cite{Jackiw:1975fn,Callan:1984sa,Kaplan:1992bt}
in 2017 \cite{Fukaya:2017tsq} with Onogi and Yamaguchi.
One year later, three mathematicians Furuta, Matuo and Yamashita joined
and we succeeded in a mathematical proof that our proposal is mathematically correct \cite{Fukaya:2019qlf}.
The reformulation is so physicist-friendly that application to lattice gauge theory
is straightforward \cite{Fukaya:2019myi} (see \cite{Onogi:2021slv} for a relation to the Berry phase).
In this conference, we had two parallel talks on an application to a curved domain-wall
fermion by Aoki \cite{AokiS} and that to the mod-two APS index \cite{Fukaya:2020tjk, Matsuki} by Matsuki.
We also refer the readers to our review paper \cite{Fukaya:2021sea} on the whole project.

\section{Massive Dirac operator index without boundary}

The key point of this work is if we can reformulate the index of the Dirac operator
in terms of massive fermions,
where either notion of chiral or zero mode is lost.
Let us start with an easier case without boundary. 

We consider a Dirac fermion with a negative mass, compared to that of the regulator.
Here we choose the Pauli-Villars regularization.
For simplicity, we couple the $SU(N)$ gauge fields to the fermion
on an even-dimensional flat Euclidean space.
The fermion determinant is expressed as
\begin{align}
  \frac{\det(D-m)}{\det(D+M)}.
\end{align}

In the large mass limit,
or simply taking the physical mass $m$ and Pauli-Villars mass $M$ the same,
let us perform an axial $U(1)$ rotation with angle $\pi$
to flip the sign of the mass.
Apparently the fermion looks decoupling from the theory,
but taking the anomaly \cite{Adler:1969gk,Bell:1969ts}
into account, we have a shift in the $\theta$ term by $\pi$
and the sign is controlled by the Atiyah-Singer index $I_{\rm AS}(D)$:
\begin{align}
  \frac{\det(D-M)}{\det(D+M)} = \frac{\det(D\textcolor{black}{+M})}{\det(D+M)}
  \times\exp\left(-i\textcolor{black}{\pi}\underbrace{\frac{1}{32\pi^2}\int d^4x FF}_{=I_{\rm AS}(D)}\right) = (-1)^{-I_{\rm AS}(D)}.
\end{align}
Even though we consider this massive fermion case,
which does not have chiral symmetry or zero modes, the index still appears.
Our proposal is then to use the massive Dirac operator to “define” the index.
We use anomaly rather than symmetry.

Specifically, multiplying $i\gamma_5$ to the numerator and denominator in the determinant
to make the operator anti-Hermitian, 
the determinant can be expressed by the products of pure imaginary eigenvalues
of massive operators:
\begin{align}
\frac{\det(D-M)}{\det(D+M)} &=\frac{\det i\gamma_5 (D-M)}{\det i\gamma_5 (D+M)} = \frac{\prod_{\lambda_{-M}}i\lambda_{-M}}{\prod_{\lambda_{+M}}i\lambda_{+M}}
= \exp\left[\frac{i\pi}{2} \left(\sum_{\lambda_{-M}}{\rm sgn}\lambda_{-M}-\sum_{\lambda_{+M}}{\rm sgn}\lambda_{+M}\right)\right].
\end{align}
Now we have shown the equality 
\begin{align}
I_{\rm AS}(D) =  \frac{1}{2}\eta(\gamma_5(D-M))^{reg.}  = -\frac{1}{2}\left[\eta(\gamma_5(D-M))-\eta(\gamma_5(D+M))\right].
\end{align}
On the right-hand side, chiral symmetry is no more essential
but it is not written by the zero modes only but also by the non-zero modes.

\section{New formulation of the index with boundary}
Next we consider the nontrivial case with boundary.
Before going into the details, let us discuss
what a more physical set-up should be.
In physics, any boundary has its ``outside''.
Topological insulator is nontrivial because its outside
is surrounded by normal insulators. 
It is more natural to regard the surface of
material as a wall between domains of different physical
profiles, rather than a boundary of a manifold.
This domain-wall should keep the
angular momentum in its normal direction,
rather than helicity of the reflecting particles.
Namely, we should consider a massive fermion system.
The boundary condition should not be put by hand
but should be automatically chosen by nature.
The edge-localized modes should play a key role.
Is there any candidate?
Yes, we have the domain-wall fermion \cite{Jackiw:1975fn,Callan:1984sa,Kaplan:1992bt}.

Let us consider a four-dimensional closed manifold,
which is extended from the original boundary,
and a massive Dirac fermion operator on it, 
\begin{align}
D+\varepsilon M, 
\end{align}
where the sign function takes $\varepsilon=-1$ in our target domain (say $x_4>0$),
and $\varepsilon=+1$, otherwise\footnote{
In \cite{Kanno:2021bze}, more general position-dependent mass is discussed.
  }.
Here we do not assume any boundary condition
on the domain-wall, expecting it dynamically given.
Unlike the standard domain-wall fermion employed
in lattice QCD simulations,
our domain-wall fermion lives in
four dimensions and the surface modes
are localized in the three dimensional wall.

Our proposal for the new expression of the index is a natural extension
of the Atiyah-Singer index in the previous section.
We find that the $\eta$ invariant
of the domain-wall Dirac operator with kink structure,
coincides with the APS index,
\begin{align}
  \label{eq:APSDW}
  I_{\rm APS}(D) =  \frac{1}{2}\eta(\gamma_5(D+\varepsilon M))^{reg.}  = \frac{1}{2}\left[\eta(\gamma_5(D+\varepsilon M))-\eta(\gamma_5(D+M))\right].
\end{align}

This equality can be shown by Fujikawa method, which consists of three steps.
1) choosing regularization: we employ the Pauli-Villars subtraction,
2) choosing the complete set to evaluate the trace: we take
the eigenmode set of free domain-wall Dirac operator squared,
and 3) perturbative evaluation.
See \cite{Fukaya:2017tsq} for the details of the computation.
Here we give two remarks about the evaluation.

First, the eigenmodes of the free domain-wall fermion,
or the solutions to 
\begin{align}
\{\gamma_5(D^{\rm free}+M\varepsilon(x_4))\}^2 \phi = \left[-\partial_\mu^2 + M^2 \textcolor{black}{-2M\gamma_4 \delta(x_4)}\right]\phi = \lambda^2 \phi 
\end{align}
are given by a direct product $\phi =\varphi_{\pm,e/o}^{\omega/{\rm edge}}(x_4) \otimes e^{i\bm{p}\cdot \bm{x}}$,
where $\bm{p}$ denotes the momentum vector in the three dimensions.
The bulk wave functions in the $x_4$ direction
\begin{align}
 \varphi^\omega_{\pm, o}(x_4)&=\frac{e^{i\omega x_4}-e^{-i\omega x_4}}{\sqrt{2\pi}},\;\;\;
  \varphi^{\omega}_{\pm,e}(x_4)=
  \frac{(i\omega\pm M)e^{i\omega |x_4|}+(i\omega\mp M)e^{-i\omega |x_4|}}{\sqrt{2\pi(\omega^2+M^2)}},
\end{align}
have the eigenvalues $\lambda^2=\bm{p}^2+\omega^2+M^2$ and
the $\pm$ eigenvalue of $\gamma_4$ indicated by the subscripts $\pm$.
Another subscript $e/o$ denotes the even/odd parity under the
reflection $x_4\to -x_4$.
In the complete set, we have the edge localized solutions with
\begin{align}
  \varphi^{\rm edge}_{-, e}(x_4)&=\sqrt{M}e^{-M|x_4|},
\end{align}
with the eigenvalue $\lambda^2=\bm{p}^2$.
These modes are chiral and massless.

The second remark is that we did not give any boundary condition by hand
but the following non-trivial boundary condition
\begin{align}
  \left[\partial_4 \mp M\varepsilon\right]\varphi^{\omega/{\rm edge}}_{\pm,e}(x_4)|_{x_4=0} = 0,\;\;\;
  \varphi^{\omega}_{\pm,o}(x_4=0)=0,
\end{align}
is automatically satisfied due to the domain-wall.
More importantly, this boundary condition preserves
the angular momentum in the $x_4$ direction but does not keep helicity.
Nature chooses the rotational symmetry, rather than chirality.

Now we can compute the $\eta$ invariant in a simple perturbative expansion.
From the bulk mode part, we obtain the curvature term but with the sign flipping:
\begin{align}
  \label{eq:bulk}
\frac{1}{2}\eta(H_{DW})^{\rm bulk}&= \frac{1}{2} \sum_{\rm bulk}(\phi^{\rm bulk})^\dagger {\rm sgn}(H_{DW})\phi^{\rm bulk}
=
\frac{1}{64\pi^2}\int d^4x \textcolor{black}{\epsilon(x_4)}\epsilon_{\mu\nu\rho\sigma}{\rm tr}_cF^{\mu\nu}F^{\rho\sigma}(x) + O(1/M),
\end{align}
where we have denoted $H_{DW}=\gamma_5(D+\varepsilon M)$.
From the edge modes, we obtain the $\eta$ invariant of the three-dimensional massless Dirac operator
on the domain-wall,
\begin{align}
  \label{eq:edge}
\frac{1}{2}\eta(H_{DW})^{\rm edge} &= \frac{1}{2}\sum_{\rm edge}\phi^{\rm edge}(x)^\dagger {\rm sgn}(H_{DW})\phi^{\rm edge}(x)
= -\frac{1}{2}\eta(iD^{\rm 3D})|_{x_4=0}.
\end{align}
Together with the Pauli-Villars contribution with $H_{PV}=\gamma_5(D+M)$,
\begin{align}
 -\frac{1}{2}\eta(H_{PV}) &=\frac{1}{64\pi^2}\int d^4x\; \epsilon_{\mu\nu\rho\sigma}{\rm tr}_cF^{\mu\nu}F^{\rho\sigma}(x) + O(1/M),
\end{align}
we finally obtain,
\begin{align}
\frac{1}{2}\eta(\gamma_5(D+\varepsilon M))^{reg.}= 
\frac{1}{32\pi^2}\int_{x_4>0} d^4x \epsilon_{\mu\nu\rho\sigma}{\rm tr}[F^{\mu\nu}F^{\rho\sigma}] \textcolor{black}{-\frac{\eta(iD^{\rm 3D})}{2}},
\end{align}
which is equal to the APS index.
The above evaluation of the bulk and edge modes separately
makes the roles of them in the anomaly inflow manifest:
the time-reversal ($T$) symmetry breaking of the edge (\ref{eq:edge})
(the $\eta$ invariant is odd in $T$) is precisely
canceled by that of the bulk modes (\ref{eq:bulk})
when they are exponentiated in the fermion partition function:
$Z_{\rm edge}Z_{\rm bulk}\propto (-1)^{-\frac{1}{2}\eta(\gamma_5(D+\varepsilon M))^{reg.}}$.

\section{Mathematical justification}

In the previous section, we have perturbatively shown on a flat Euclidean space
that the $\eta$ invariant of the domain-wall Dirac operator equals to the APS index
of a domain with negative mass, where the APS boundary condition is imposed on the
location of the domain-wall, in spite of the fact that we did not impose any
boundary condition in the former set up.
As the two formulations are given on different manifolds,
the reader may wonder if the equivalence is just a coincidence.

Since the problem is nontrivial not only in physics but also in mathematics,
our collaboration invited three mathematicians and
the interdisciplinary collaboration went successful
in giving a general proof for the equivalence.
In \cite{Fukaya:2019qlf} we gave a proof that for any APS index of a Dirac operator
on a manifold $X_+$ with boundary $Y$, there exists
a massive domain-wall fermion Dirac operator on a closed manifold $X$
extended from $X_+$, and its $\eta$ invariant equals to the original index.

Here we just give a sketch of the proof.
We introduce an extra dimension to consider $X \times \mathbb{R}$, and
the following Dirac operator on it,
\begin{align}
 D^{\rm 5D} = \left(\begin{array}{cc}
0 & \partial_5 + \gamma_5 (D^{\rm 4D} + m(x_4,x_5))\\
-\partial_5 + \gamma_5 (D^{\rm 4D} + m(x_4,x_5)) & 0
\end{array}\right),
\end{align}
where the mass term is given negative in a positive region of $x_4$ and $x_5$ and positive otherwise:
\begin{align}
m(x_4,x_5) = \left\{
\begin{array}{cc}
-M & \mbox{for}\; x_4>0\; \&\; x_5>0\\
0 & \mbox{for}\; x_4=0\; \&\; x_5=0\\
M_2 & \mbox{otherwise}\\
\end{array}\right.
\end{align}
Here the gauge fields are copied in the $x_5$ direction
and we set $A_5=0$.
Then we evaluate the index of this Dirac operator
in two different ways.
With the so-called localization technique \cite{Witten:1982im,FurutaIndex}, which focuses on
the edge-localized modes on the domain-wall, 
we can show that the index is equal to the original APS index,
together with an equality between the index and 
that with an half-infinite cylinder.
The index can be also computed by counting
the zero-crossings of the eigenvalues of
the Dirac operator in the 4 dimensions,
which leads to our new proposal, the 
$\eta$ invariant of domain-wall Dirac operators.
The equality (\ref{eq:APSDW}) always holds since the left and right-hand sides
are just two different evaluations of the same quantity.

\section{APS index on a lattice}

Finally let us discuss the lattice version of the APS index.

In the standard lattice gauge theory on a square hyper-cubic
and periodic lattices, it is well-known that
the overlap fermion action \cite{Neuberger:1997fp}
$S = \sum_x \bar{q}(x) D_{ov}q(x)$ is invariant under
a modified chiral transformation \cite{Ginsparg:1981bj, Luscher:1998pqa}:
\begin{align}
q \to e^{i\alpha\gamma_5(1-aD_{ov})}q,\;\;\;\bar{q}  \to \bar{q}e^{i\alpha\gamma_5}.
\end{align}
But the fermion measure transforms as
\begin{align}
Dq\bar{q} \to \exp\left[2i\alpha {\rm Tr}(\gamma_5+\gamma_5(1-aD_{ov}))/2\right]D q\bar{q},
\end{align}
which reproduces the axial $U(1)$ anomaly.
Moreover, the AS index can be defined as
\begin{align}
  \label{eq:latticeAS}
I_{\rm AS}(D_{ov})={\rm Tr}\gamma_5\left(1-\frac{aD_{ov}}{2}\right),
\end{align}
and it can reach the continuum value even at a finite lattice spacing \cite{Hasenfratz:1993sp},
when the gauge link variables are smooth enough.

But how to formulate the APS index has not been known.
In continuum theory, the APS boundary condition is imposed by separating the
normal and tangent parts of the Dirac operator to the boundary
and requiring the positive components of the tangent Dirac operator
of fermion fields to vanish at the boundary.
On the lattice, however, such a separation is difficult
as is clear from the explicit form of the overlap Dirac operator,
\begin{align}
  \label{eq:Dov}
D_{ov} = \frac{1}{a}\left(1+\gamma_5\frac{H_W}{\sqrt{H_W^2}}\right),
\end{align}
where $H_W=\gamma_5(D_W-1/a)$ is the Hermitian Wilson Dirac operator.
Even if we managed to impose a lattice version of the
APS boundary condition, the Ginsparg-Wilson relation or
the definition of the index would  no longer been guaranteed.

In fact, an alternative direction is hidden in the index of the overlap Dirac operator.
If we substitute the explicit form (\ref{eq:Dov}) into the
definition of the index (\ref{eq:latticeAS}), we end up with
the $\eta$ invariant of a negatively very massive Wilson Dirac operator,
\begin{align}
I_{\rm AS}(D_{ov})= - \frac{1}{2}{\rm Tr}\frac{H_W}{\sqrt{H_W^2}} = - \frac{1}{2}\eta(\gamma_5(D_W-1/a)).
\end{align}
Interestingly, the lattice index ``knew'' 
1) the AS index can be written by the $\eta$ invariant of a massive Dirac operator\footnote{
  This fact was known, for example, in \cite{Itoh:1987iy, Adams:1998eg}
  but its mathematical importance was not discussed, as far as we know.
  },
and 2) chiral symmetry is not important: the Wilson Dirac operator is enough to define it\footnote{
The issue was recently revisited by some of mathematicians \cite{Yamashita:2020nkf,Kubota:2020tpr}.
}.

The situation can be summarized in two tables.
With the conventional massless Dirac operator, we need a significant effort to formulate the index theorems.
For the APS index we need the unphysical boundary condition.
For the lattice Atiyah-Singer index, we need the overlap fermion.
The lattice APS is even not known yet.
However, the $\eta$ invariant with massive Dirac operator on a closed manifold
gives a united treatment of index theorems which is easy to handle.
The APS in continuum theory is given by a kink structure in the mass term
and the lattice AS is given by Wilson Dirac operator.
Now we can easily speculate that the lattice APS index
must be given by the eta-invariant of the Wilson Dirac operator with a domain-wall mass
$-\frac{1}{2}\eta(\gamma_5(D_W-\varepsilon M))$.

\begin{table}[hbt]
  \caption{The standard formulation of the index with massless Dirac operator}
      \label{tab:massless}
  \centering
  \begin{tabular}{|c|c|c|}
    \hline
    & continuum & lattice\\\hline
    AS & ${\rm Tr}\gamma_5e^{D^2/M^2}$ & ${\rm Tr}\gamma_5(1-aD_{ov}/2)$\\\hline
    APS  & ${\rm Tr}\gamma_5e^{D^2/M^2}$ w/ APS b.c. & not known.\\\hline
  \end{tabular}
  \vspace{1cm}
  \caption{The $\eta$ invariant of massive Dirac operator}
  \centering
  \begin{tabular}{|c|c|c|}
    \hline
    & continuum & lattice\\\hline
    AS & $-\frac{1}{2}\eta(\gamma_5(D- M))^{reg.}$ & $-\frac{1}{2}\eta(\gamma_5(D_W-M))$\\\hline
    APS  & $-\frac{1}{2}\eta(\gamma_5(D-\varepsilon M))^{reg.}$ & $-\frac{1}{2}\eta(\gamma_5(D_W-\varepsilon M))$ \\\hline
  \end{tabular}
  \label{tab:massive}
\end{table}

  In \cite{Fukaya:2019myi}, we have perturbatively shown on a four-dimensional Euclidean flat lattice that
  $-\frac{1}{2}\eta(\gamma_5(D_W-\varepsilon M))$ in the classical continuum limit is
  \begin{align}
    \label{eq:latAPS}
-\frac{1}{2}\eta(\gamma_5(D_W-\varepsilon M))=& \displaystyle{\frac{1}{32\pi^2} \int_{0<x_4<L_4} d^4x \epsilon^{\mu\nu\rho\sigma} {\rm tr} F_{\mu\nu}F_{\rho\sigma}}
  -\frac{1}{2}\eta(iD^{3\mathrm{D}})|_{x_4=0}+\frac{1}{2}\eta(iD^{3\mathrm{D}})|_{x_4=L_4},
\end{align}
  which coincides with the APS index on $T^3\times I$ with $I=[0,L_4]$.
Here we have put two domain-walls at $x_4=a/2$ and $x_4=L_4-a/2$ and set $M=1/a$.
Since the left-hand side of (\ref{eq:latAPS}) is always an integer,
we expect that this definition is non-perturbatively valid.

\section{Summary}

In this work, we have shown that the massive domain-wall fermion
is physicist-friendly: the APS index can be formulated (even on a lattice)
without any unphysical boundary conditions.
Moreover, it is mathematically rich:
the $\eta$ invariant of the massive Dirac operator
on a closed manifold unifies the index theorems.

The author thanks M.~Furuta, N.~Kawai, S.~Matsuo, Y.~Matsuki, M.~Mori, K.~Nakayama,
T.~Onogi, S.~Yamaguchi and M.~Yamashita for the fascinating collaborations.
This work was supported in part by JSPS KAKENHI Grant Number
JP18H01216 and JP18H04484.


\end{document}